\documentclass{emulateapj}

\accepted{1 October 2004}

\lefthead{Wada, Tomisaka}
\righthead{Observing the obscuring torus}

\newcommand{\bea}{\begin{eqnarray} }
\newcommand{\eea}{\end{eqnarray}}

\newcommand{\twco}{$^{12}$CO }
\newcommand{\thco}{$^{13}$CO }

\begin{document}

\title{Molecular gas structure around an AGN with nuclear starburst: 
3-D Non-LTE calculations of CO lines}

\author{Keiichi WADA$^1$}
\author{Kohji TOMISAKA$^1$}
\affil{National Astronomical Observatory of Japan, Mitaka, Tokyo
181-8588, Japan\\
E-mail: wada.keiichi@nao.ac.jp}
\altaffiltext{1}{Department of Astronomical Science, The Graduate University for Advanced Studies, Osawa 2-21-1, Mitaka, Tokyo 181-8588, Japan.}




\begin{abstract}
We have performed three-dimensional, non-LTE (non-Local Thermodynamic
Equilibrium) radiative transfer calculations for $^{12}$CO and
$^{13}$CO lines, applying them to our high-resolution hydrodynamic
models of the `torus' around a supermassive black hole in an active
galactic nucleus.  The hydrodynamic simulations reveal inhomogeneous
and turbulent gas structure on a sub-pc scale in a circum-nuclear
starburst region.  Thick disks interlaced with filaments, clumps and
holes are naturally formed due to the interplay among energy feedback from
supernovae, self-gravity of the gas, galactic rotation, and radiative
cooling.  The intensity maps of the molecular lines for the
circum-nuclear disks show a clumpy structure reflecting the intrinsic
inhomogeneity and turbulent motion of the gas disk. The fine structure
of the `torus' could be resolved in the nearby active galaxies using
the Atacama Large Millimeter/submillimeter Array. We also found that
the CO-to-H$_2$ conversion factor (X-factor) is not uniformly
distributed in the central 100 pc region. The X-factor derived for
$^{12}$CO ($J=$1-0) intensity depends strongly on the intensity,
whereas the X for $^{12}$CO ($J=$3-2) is nearly constant over two
orders of magnitude of the intensity. The suggested conversion factor
for the molecular gas mass is $X_{{\rm CO} (J=3-2)} \sim 0.27 \times
10^{20}$ cm$^{-2}$ (K km s$^{-1}$)$^{-1}$.  The line-ratio of high-J
transitions of CO is not uniformly distributed in the nuclear disk,
and the apparent ratio depends on the beam sizes. `Observed' \twco
(J=4-3)/\twco (J=2-1) can differ from the intrinsic ratio by maximally
20\%.

\end{abstract}




%


%
\section{INTRODUCTION}
%
The galactic centers of spiral galaxies are important regions for
the evolution of galaxies. It has been widely believed that most `large' galaxies 
harbor supermassive black holes (SMBHs) at their centers, and also that the observed
scaling relation between masses of black holes and the galactic bulges
suggests that they have evolved together during the hierarchical formation of
galaxies \citep[e.g.][]{kauf00}. The galactic centers are
often very active. Nuclear starbursts and various kinds of 
active galactic nuclei (AGNs) are observed in nearby galaxies and
also in galaxies at high redshifts.  These active phenomena must be related with
the evolution of the SMBHs and galaxies, as well as the intergalactic
medium through galactic winds and radiative feedback, especially at high
redshifts.  All these phenomena are related to
the structure and dynamics of the
interstellar medium (ISM) in the central 100 pc regions of galaxies.
For example, AGN activities are supported by gas fueling, whose mechanisms
remain unclear, especially in the inner 100 pc from the central engine (see Wada 2004,
and references therein). We have not yet determined what triggers the nuclear
starbursts that are connected to the formation of stellar cores
and the SMBH \citep[e.g.][]{norm88}.
However, in the next decade, we may expect drastic change in
our observational knowledge of the ISM in the circum-nuclear region of
galaxies. The $\sim$ 0.01'' spatial resolution with the next generation millimeter and submillimeter interferometer, ALMA (the Atacama Large Millimeter Array) 
may reveal a sub-pc structure of the ISM.
We now need to improve our theoretical modeling of the ISM in the
galactic central regions
in order to make direct comparison with future observations.

The ISM in galaxies is inhomogeneous and 
multi-phase, and this is especially important
in studying the gas dynamics in circum-nuclear regions, 
because the typical scale of the inhomogeneity of the ISM ($\sim 10$ pc) is not
negligible compared to the size of the system ($\sim 100$ pc).
The isothermal approximation (e.g. the sound velocity $\sim $ 10 km s$^{-1}$),
which has often been used for modeling gas dynamics on a galactic scale,
is not adequate for the nuclear gas dynamics,
particularly when energy and radiative feedback from stars is taken into account.
The multi-phase nature is also important for comparing 
numerical models with observations. Radiative transfer calculations, 
for continuum and various lines emitted from the AGN regions,
require realistic density, temperature
and velocity distributions of the ISM as input.
In order to achieve realistic numerical models of 
the ISM, sub-pc spatial resolution for diffuse gases as well as for dense gases and
resolving strong shocks is also essential.
We should also take into account the self-gravity 
and three-dimensional dynamics of the ISM.

Instead of adopting phenomenological approaches for the multi-phase ISM,
in which many free parameters and assumptions are unavoidable \citep[e.g.][]{comb85},\citet{wad01b, wad02} obtain global, inhomogeneous models of the ISM with a sub-pc scale resolution,
by solving the governing basic equations of the ISM with a high-accuracy Euler 
mesh code. In the present paper,  using these results (three-dimensional density, 
temperature and velocity field data),
we have made the first attempt to derive the molecular line intensities 
emitted from the nuclear starburst region with a supermassive black hole
applying our non-LTE (non Local Thermodynamic Equilibrium), 
fully three-dimensional radiative transfer calculations.

These kinds of `synthesized observations' based on radiative transfer
calculations has recently been applied to turbulent molecular clouds \citep{pad98,oss02, oss02b}. \citet{pad98} performed non-LTE radiative transfer calculations 
of CO and 
 CS lines using a Monte-Carlo integration scheme \citep{juv97} 
for numerical results of the magneto-hydrodynamic simulations of turbulent
clouds. They found that the synthesized maps and spectra bear resemblance to
real molecular clouds in many respects. \citet{oss02} and \citet{oss02b} made  similar analysis of turbulent clouds, but they assumed
a local LVG (Large Velocity Gradient) approximation and isotropic radiation field. 
Ossenkopft et al. claimed that their simplified analysis is accurate enough
for comparison between the observations of molecular clouds and 
the numerical models of turbulent clouds.
These studies focused on the molecular clouds in disks of spiral galaxies,
like our Galaxy, therefore we cannot simply apply their results to
the interstellar medium in the nuclear starburst with a supermassive
black hole. Apparently we should not assume isothermal, isotropic turbulence
in the galactic nuclear region. Since the local turbulent motion is of the
same order as the global rotation \citep{wad02}, 
a fully consistent treatment of the radiative transfer is necessary.

In \S 2, we describe the numerical method and models. In \S 3, the typical
structure of the ISM in the circum-nuclear region
obtained by hydrodynamic simulations is briefly described. Then, 
$^{12}$CO and $^{13}$CO line intensity calculations are shown.
 The CO-to-H$_2$ conversion factor, the so-called ``X'' factor and the line-ratios are also discussed. A brief summary and discussion are presented
in \S 4.

%
\section{Numerical Method and Models}
%
\subsection{Hydrodynamic models}
The numerical methods for obtaining three-dimensional density, temperature 
and velocity fields of the gas around a SMBH, for which the line-transfer and excitation of molecules are calculated, are the same as those described 
in \citet{wad02}. Details of the two-dimensional version of the numerical scheme 
is described in \citet{wad01b}. In addition to the model described in \citet{wad02},
we also performed a model with a moderate supernova rate and a less massive black hole (see below). Here, we briefly summarize them.
\setcounter{footnote}{0} We solve mass, momentum, and energy conservation equations
with the Poisson equation  numerically
in 3-D to simulate the evolution of a rotating ISM in 
a fixed gravitational potential. Here the external potential force in 
the momentum conservation equation is
$\nabla \Phi_{\rm ext} + \nabla \Phi_{\rm BH}$, 
where 
the time-independent
external potential is $\Phi_{\rm ext} \equiv -(27/4)^{1/2}v_c^2/(r^2+
a^2)^{1/2}$ with a core radius of $a = 10$ pc and
a maximum rotational velocity of $v_c = 100$ km s$^{-1}$. 
The central BH potential is
$\Phi_{\rm BH} \equiv -GM_{\rm BH}/(r^2 + b^2)^{1/2}$ with $b=1$ pc.
The mass of the BH is assumed to $10^8 M_\odot$ (model A) and $10^7 M_\odot$ (model B).  Model A is the same model studied in \citet{wad02}.
We also assume a cooling function $\Lambda(T_g) $ $(5 K < T_g < 10^8
{\rm K})$ with Solar metallicity and heating due to
photoelectric heating, $\Gamma_{\rm UV}$ and due to energy feedback
from SNe, $\Gamma_{\rm SN}$.  We assume a uniform UV radiation field,
which is ten times larger than the local UV field.
We assume an equation of state for an ideal gas with a ratio of
specific heat $\gamma = 5/3$. 

The hydrodynamic part of the basic equations is solved by AUSM
 (Advection Upstream Splitting Method) (Liou \& Steffen 1993).
We use $256^2 \times 128$ (model A) or $256^3$ (model B) 
Cartesian grid points covering a $64^2\times 32$ pc$^3$ (model A) or 
$64^3$ pc$^3$ (model B) region around the galactic center (the
spatial resolution is 0.25 pc).  The Poisson equation is solved 
using the fast Fourier transform and the convolution method.
The initial condition is an axisymmetric and rotationally
supported thin disk with a uniform radial density profile (the scale height in
z-direction is 2.5 pc) and a total gas mass of $M_g = 5\times 10^7 M_\odot$ (Model A) and 
$M_g = 10^7 M_\odot$ (Model B).
Random density and temperature fluctuations, which are 
less than 1 \% of the unperturbed values, 
are added to the initial disk.

Supernova (SN) explosions are assumed to occur at random positions
on the disk plane.
The average SN
rate is $\sim$ 0.8 yr$^{-1}$ (model A) and 0.08 yr$^{-1}$ (model B).
The energy of $10^{51}$ ergs is instantaneously
injected for each supernova into a single cell, as a form of thermal energy. 
The three-dimensional evolution of blast waves driven by the SNe in an inhomogeneous and non-stationary medium with a global
rotation is followed explicitly, taking into account the radiative cooling.
Therefore the evolution of the SN remnants, e.g. their duration and structure,
 depends on the gas density distribution around the SNe, 
and we do not need to assume free parameters, such as the heating efficiency.


\subsection{Line radiative transfer calculations}
Next, we try to `observe' the star-forming disk around a 
SMBH, calculated separately using the hydrodynamic method described in
\S 2.1,  with molecular lines.
A snapshot of density, temperature and velocity fields of $128^2 \times 64^2$ (model A) or $128^3$ (model B) or cells 
[i.e. each cell is (0.5 pc)$^3$], averaged  from the original 
hydrodynamics simulations with $256^3$ or $256^2 \times 128$, 
cells is used to obtain  \twco and \thco intensities.
We have performed three-dimensional, fully non-LTE calculations 
to obtain CO line intensity distribution. 
We use the density, temperature and velocity data in a $256^3$ data cube
from a snapshot of a  three-dimensional hydrodynamical calculation, and 
the accelerated Monte-Carlo method is applied. 
We have developed 
a three-dimensional code, based on Hogerheijde \& van der Tak (2000),
and this is optimized to our vector-parallel supercomputer.
Here we briefly describe the method.
The equation of radiative transfer is
\bea
\frac{dI_\nu}{d\tau_\nu} = -I_\nu + S_\nu,
\label{eq: 1}
\eea
where $I_\nu$ is the intensity at frequency $\nu$ along a particular
line of sight described $d\bf{s}$, the source function $S_\nu$ is 
defined as $S_\nu \equiv j_\nu/\alpha_\nu$ with emission and 
absorption coefficients $j_\nu$ and $\alpha_\nu$. The optical depth is
defined as $d \tau \equiv \alpha_\nu d\bf{s}$.
The coefficients $j_\nu$ and $\alpha_\nu$ for a spectral line are determined by
absorption and emission between levels $u$ and $l$ with number density
$n_u$ and $n_l$:
\bea
j_\nu^{ul} &=& \frac{h\nu_0}{4 \pi} n_u A_{ul} \phi(\nu), \\
\alpha_\nu^{ul} &=& \frac{h\nu_0}{4 \pi} (n_lB_{lu} - n_u B_{ul}) \phi(\nu),
\label{eq: 2}
\eea
where $\phi(\nu)$ is a line-profile function peaked around the frequency $\nu_0 = (E_u -E_l)/h$, and $A_{ul},B_{lu}$, and $B_{ul}$ are the Einstein probability coefficients.
The line profile $\phi(\nu)$ is given by
\bea
\phi(\nu) = \frac{1}{\sqrt{\pi}\sigma\nu_0} \exp\left[-\frac{c^2 (\nu-\nu_0)^2}{\nu_0^2 \sigma^2}\right].
\label{eq: 3}
\eea
The level populations are determined by the equation of statistical equilibrium, i.e.
\bea
n_l [ \Sigma_{ k < l} A_{lk} + \Sigma_{k\ne l} (B_{lk} J_\nu + C_{lk})] = \\ \nonumber 
\Sigma_{ k > l} n_k A_{kl} + \Sigma_{k\ne l} n_k(B_{kl} J_\nu + C_{kl}),
\label{eq: 4}
\eea
where $J_\nu$ is the mean intensity of the radiation field,
\bea
J_\nu \equiv \frac{1}{4\pi} \int I_\nu d\Omega,
\label{eq: 5}
\eea
with $\nu \equiv |E_k - E_l|/h$.
Equations (\ref{eq: 1}), (\ref{eq: 4}) and (\ref{eq: 5}) 
are solved iteratively.
The integral in eq. (\ref{eq: 5}) is evaluated using the Monte Carlo approach.
Instead of using the original Monte Carlo approach \citep{bern79}, 
here we used the implementation developed by \citet{hoge00},
in which the integration is performed along the rays incident on each cell from 
infinity. This approach has the advantage of convergence of the radiation field and
level populations.

For each grid cell, $N_{\rm Ray}$ rays are calculated.
First, we choose the wavelength of photons $\nu$ incoming to the cell
at ${\bf x}$ according to equation (\ref{eq: 3}). The central frequency of these
photons is taken to be equal to $\nu_L\left[1+{\bf n}\cdot {\bf v}({\bf x})/c\right]$,
where $\nu_L$, ${\bf n}$, and ${\bf v}$ represent the central frequency
of this transition in the laboratory frame, randomly chosen normal vector
of the ray from ${\bf x}$ and the local velocity of the gas at  $\bf{x}$.
For these photons, the absorption and emission due to the grid cell at
${\bf x}'$ on the path are calculated using equations (2) and (3).
In equation (\ref{eq: 3}), we put $\nu_0=\nu_L\left[1+{\bf n}\cdot {\bf v}({\bf x}')/c\right]$
for the central frequency.  Then, we integrate equation (\ref{eq: 1}) from the boundary,
where the cosmic microwave background radiation is assumed,
and obtain $I_\nu({\bf x})$.  After averaging the specific intensities
of $N_{\rm Ray}$ rays as equation (\ref{eq: 5}) by adding respective intensities,
we finally solve the rate equation and obtain the level populations for
the cell at ${\bf x}$.  This procedure continues till the level populations
of each cell converge.

We solved ten excitation levels of $^{12}$CO and $^{13}$CO, 
and $N_{\rm Ray} = 200$ for each cell is used. In most cases, the radiation field, 
the level populations, and the final intensity distribution of the lower-level ($J < 6$) molecular lines are converged in about 20 iterations. 
The collision coefficients for $^{12}$CO and $^{13}$CO are taken from
\citet{gre76} and \citet{gre78}. Ten energy levels of these molecules 
are included in the calculation.
We have assumed constant abundances of
$5 \times 10^{-5}$ for $^{12}$CO and $10^{-6}$ for $^{13}$CO.
The `micro-turbulence', $\sigma$ in eq.(\ref{eq: 3}) is assumed 1 km s$^{-1}$.

Computational time for one typical run of these radiative transfer calculations
is typically about 4 hours, using 32 processors of 
a vector-parallel supercomputer Fujitsu VPP5000 
(the peak performance is 9.6 Gflops/processor).

%
\section{Numerical Results}
%
\subsection{Input models}
Recent high resolution, 3-D  hydrodynamical simulations revealed that 
a typical three-dimensional density structure of the ISM in the central 100 pc 
is characterized by a `tangled' network of filaments and 
dense clumps \citep{wada01}.
These filaments are a natural consequence of 
the gravitational and thermal instabilities, tidal interaction between
dense clumps, and local and global shear motions.
The probability distribution function of gas density in an equilibrium 
state is well represented by a single log-normal function 
 (Wada \& Norman 2001; Wada 2001). In other words, this reflects 
the fact that the denser media occupy smaller volumes, and 
the log-normal function implies that the inhomogeneous structure 
is a consequence of highly non-linear ways. This nature of the 
density field would be related to the empirical relation 
between the global star formation rate and the average surface density
of the ISM \citep{elm02}.
During the development of these inhomogeneous structures,
a small part of the galactic rotational energy is converted to 
maintain a steady turbulent motion of the ISM 
[see also Wada, Meurer, \& Norman (2002) for two-dimensional cases].
The ISM around a SMBH ($10^{6-9} M_\odot$) on a
10-100 pc scale is basically the same structure as mentioned above.
However, the gravitational field of the massive black hole 
affects the scale height of the disk. Energy feedback from 
supernova explosions, which is expected in the clumpy medium, also
determines the global structure of the nuclear gas disk. 
Therefore, the scale height of the thick disk is determined from a balance between supernova heating and turbulent energy dissipation due to the radiative cooling in a black hole and galactic gravitational field \citep{wad02}.


\begin{figure}[p]
\centering
\plotone{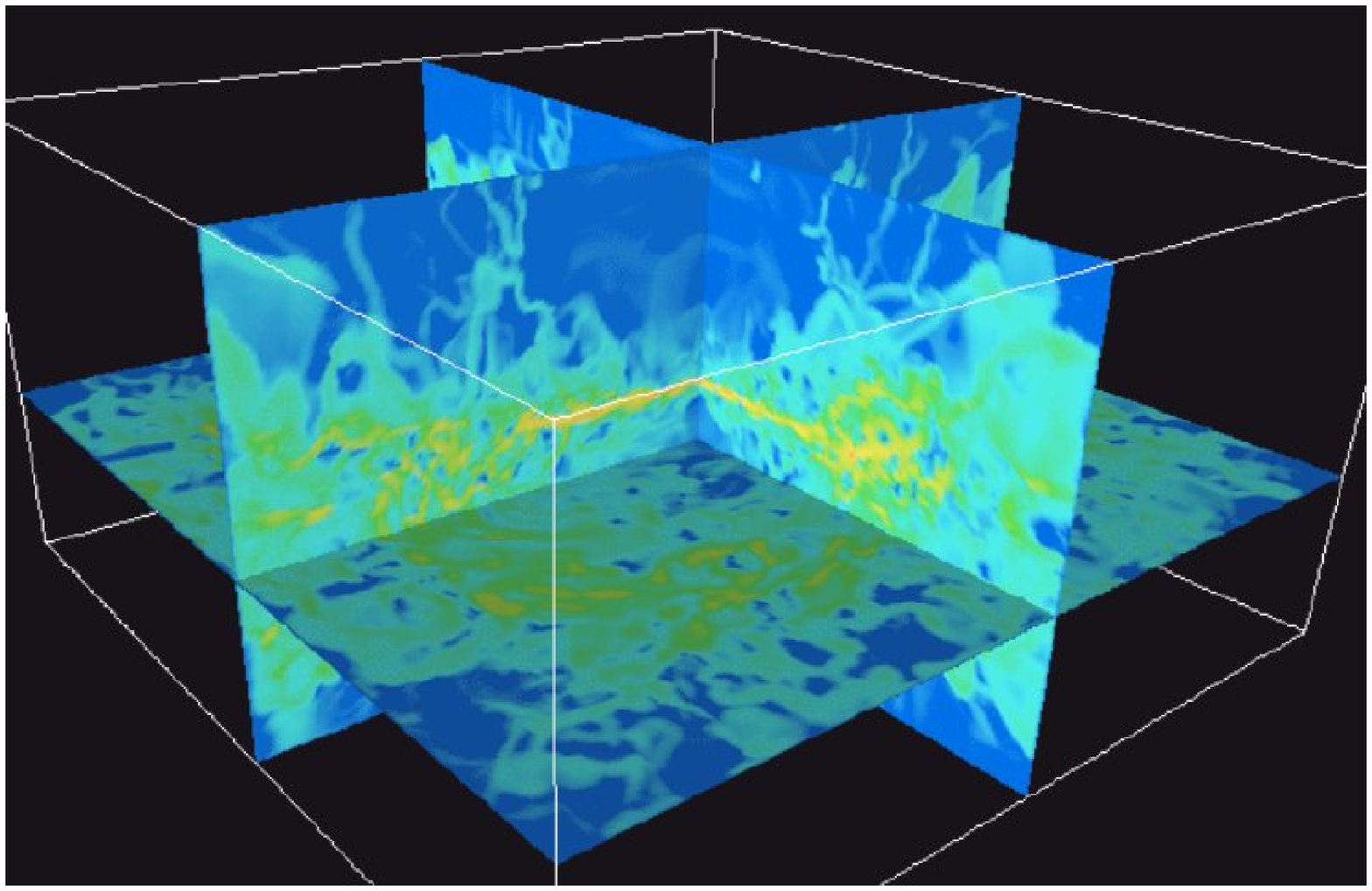}
\caption{Surface sections of the density distribution of the gas around 
a SMBH (model A).  x-y, y-z, and z-x planes are shown. 
False color is coded according to the local volume density. 
The yellow regions represent $n > 10^3$ cm$^{-3}$. The box size is 64 pc $\times$ 64 pc $\times$ 32 pc.}
\label{wada_fig: f1}
\end{figure}

Figure \ref{wada_fig: f1} is density distribution of model A ($M_{\rm BH} 
= 10^8 M_\odot$ with a relatively high star formation rate $\sim 0.8 M_\sun$ yr$^{-1}$), and it is clear that
the gas around the central massive black hole is filamentary and clumpy, 
and the scale height is larger in the outer region. As pointed out in Wada \& Norman (2002), the internal inhomogeneous structure of the 'torus' is not steady, but rather seemingly turbulent, although it globally rotates around the center.
The global flared disk shape does not change significantly during many rotational
periods. 
The turbulent motion is enhanced by supernova explosions which are 
assumed to randomly occur in the disk,
and their blast waves blow the gas up from the 
disk plane.

\begin{figure}[t]
\centering
\plotone{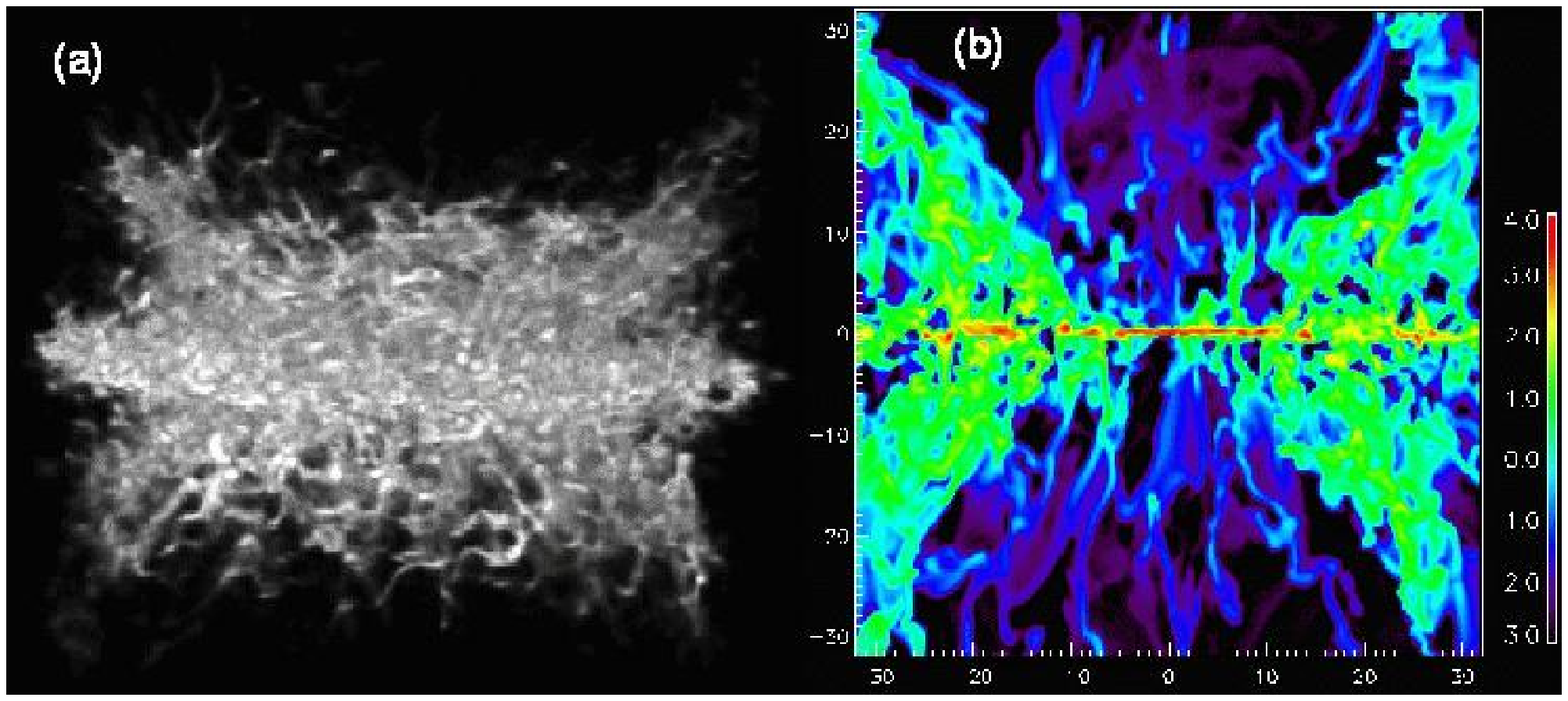}
\caption{``Tangled network'' of the gas around a SMBH with the nuclear 
starburst of Model B. (a) Volume rendering representation of the density 
field. The gray-scale represents `relative column density' 
or `relative opacity' distribution, 
but its absolute value does not correspond to any physical scales.
The box size is 70 pc across. (b) Surface section of the density field of
Model B. The color is log-scaled volume density ($M_\odot$ pc$^{-3}$).
 $256^3$ grid cells are used for a ($64$ pc)$^3$ computational box.}
\label{wada_fig: f2}
\end{figure}

Figure \ref{wada_fig: f2} shows the density structure of model B, 
in which a less active supernova rate (0.08 yr$^{-1}$) 
and a less massive black hole ($10^7 M_\odot$) are assumed. 
The structure is similar to those seen in Figure \ref{wada_fig: f1}.

\subsection{Molecular line-transfer calculations}

\begin{figure}[p]
\centering
\plotone{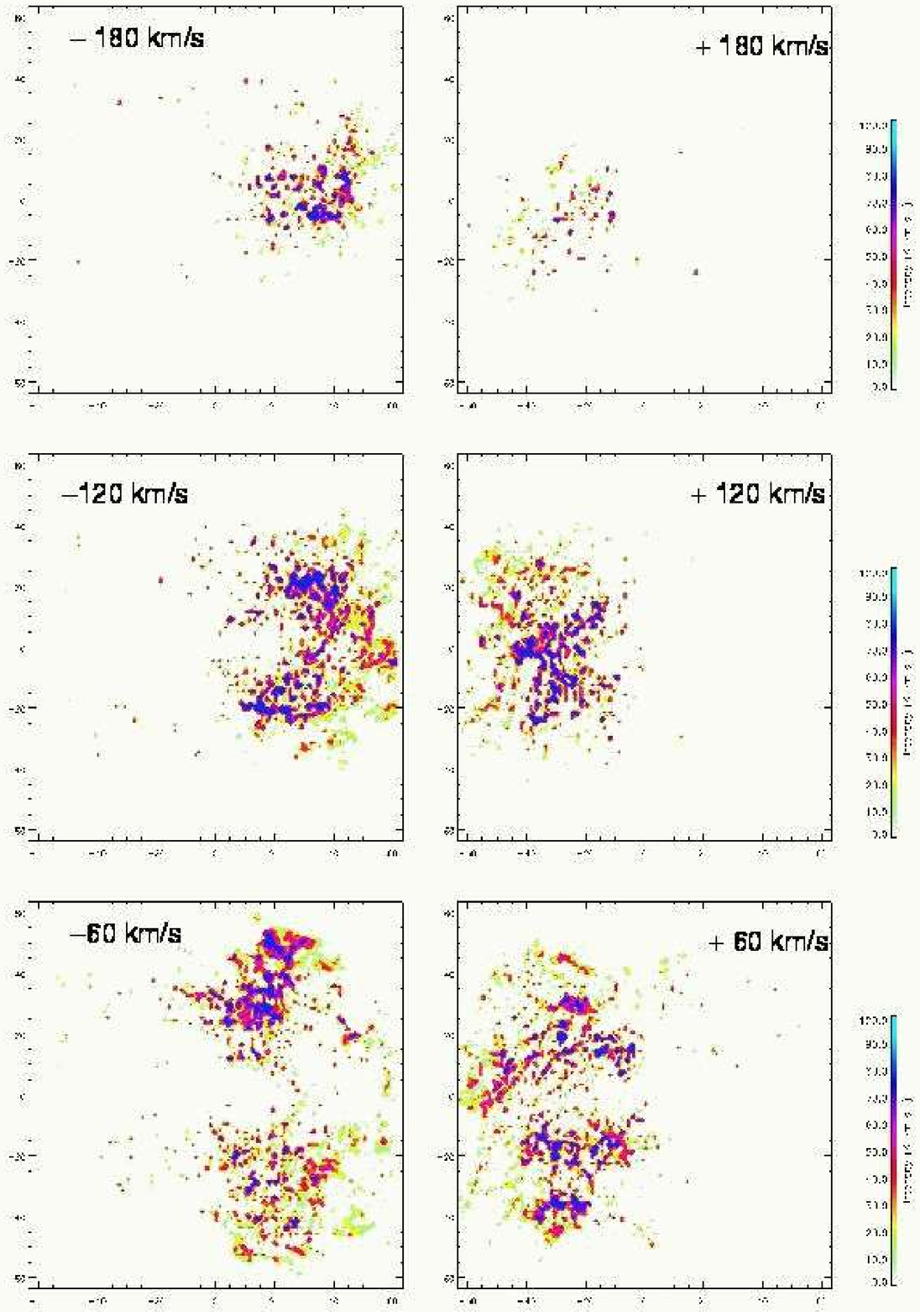}
\caption{Channel maps of \twco $(J=2-1)$ for a snapshot of model A (Figure 1). 
Six velocity channels from -180 km s$^{-1}$ to
$+180$ km s$^{-1}$ are picked up from 100 channels in $\pm$ 500 km s$^{-1}$.
A viewing angle of the torus 45 degrees from the rotational axis 
is assumed. }
\label{wada_fig: f3}
\end{figure}

Since most parts of the gas mass in the torus are in cold ($T < 100$ K) and 
dense ($n > 10^2$ cm$^{-3}$) phases, we expect that 
molecular lines, such as \twco ($J=$1-0, 2-1, ...), are good tracers to observe the molecular torus.
Here we demonstrate how the dense clumpy `torus' is {\it observed}
by the $^{12}$CO lines, using 
three-dimensional radiative transfer calculations, without assuming 
the LTE described in \S 2. 
Figure \ref{wada_fig: f3} shows channel maps of \twco ($J=2-1$)  for 
the torus shown in Figure 1 
with a viewing angle of 45 degrees from the rotational axis.
If we obtain channel maps like Figure 2 by future radio observations, 
we can confirm the ISM in the circum-nuclear region is globally rotating
at $\sim 200$ km s$^{-1}$, 
and the internal velocity dispersion for the cold and dense gas
is as large as $\sim 50$ km s$^{-1}$. 
The large random motion inside the torus 
suggests that the torus is geometrically thick, provided that 
the scale height of the torus is determined 
by a balance between vertical components of the centrifugal force 
due to the central black hole and vertical velocity dispersion 
\citep[see][]{wad02}.

\begin{figure}[t]
\centering
\plotone{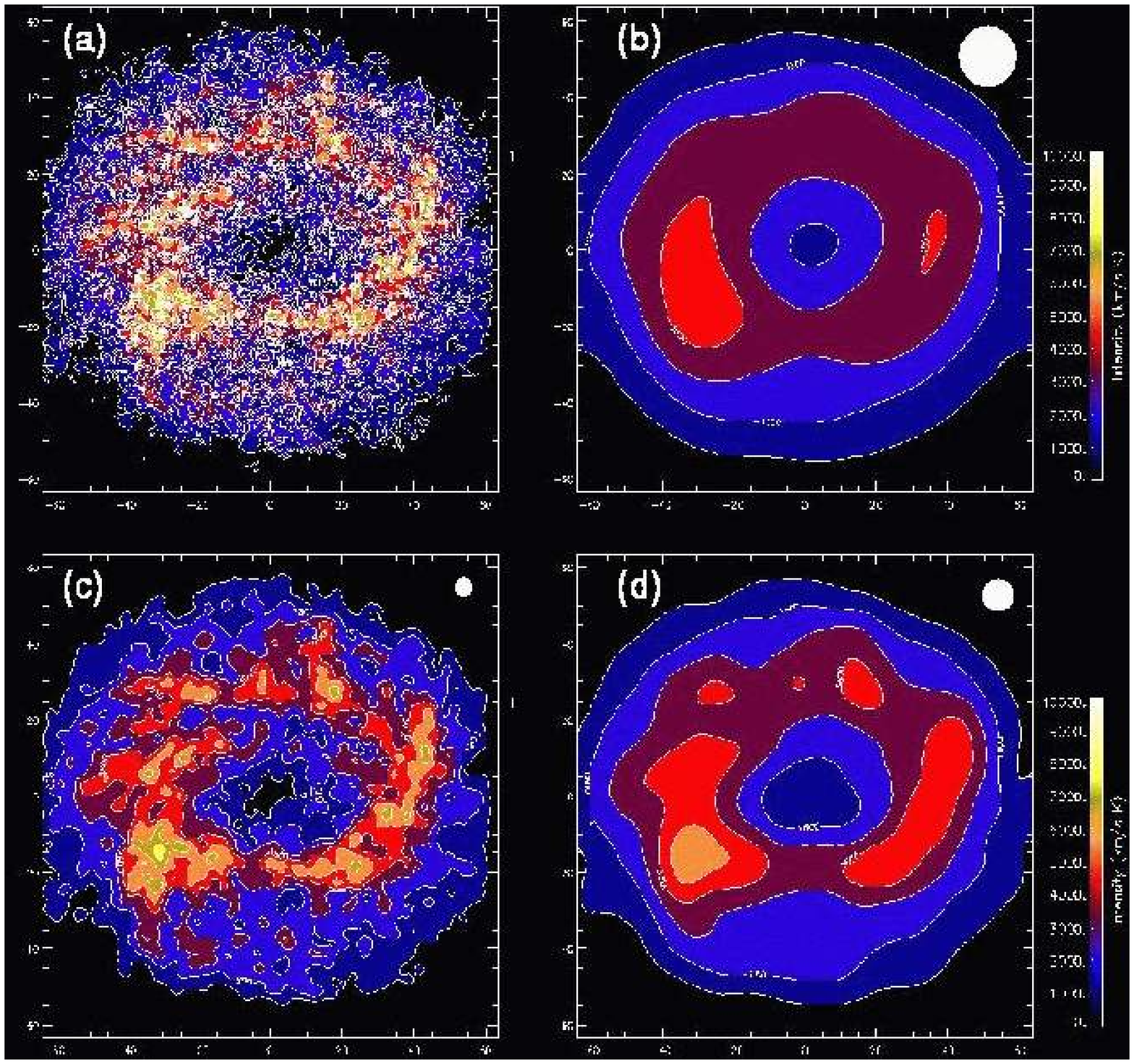}
\caption{Integrated intensity maps of \twco ($J=$2-1) calculated from
the density, temperature and velocity field of model A shown in Figure \ref{wada_fig: f2}. 
Four different `beam sizes'(shown in white filled circles)
 are assumed to convolve the original map. The unit of axes is the number of grid points (i.e. 0.5 pc/grid)}
\label{wada_fig: f4}
\end{figure}

Figure \ref{wada_fig: f4} shows integrated intensity maps for model A in Figure \ref{wada_fig: f1}. 
We convolved the maps with four different `beam' sizes to see how the
difference affects the apparent structure 
inhomogeneity of the integrated maps. As is clearly seen in Figure 3b,  
the clumpy structure of the torus cannot be resolved with a beam size of 
about 1/10 of the disk size, although the torus-like structure can be still observed.
Suppose this torus is located in the Virgo cluster, 0.01 arcsec angular resolution,
which is expected to be attained by the ALMA,  
approximately corresponds to the beam size in Figure 3c.
Therefore, the clumpy structure of the molecular tori will easily be detected. 
On the other hand, the one arcsec beam, 
which is typically achieved by present-day radio interferometers, 
such as the Nobeyama Millimeter Array, is almost comparable to the torus size,
therefore even the central hole of the torus could not be resolved, even if 
it in fact presents. 

\begin{figure}[t]
\centering
\plotone{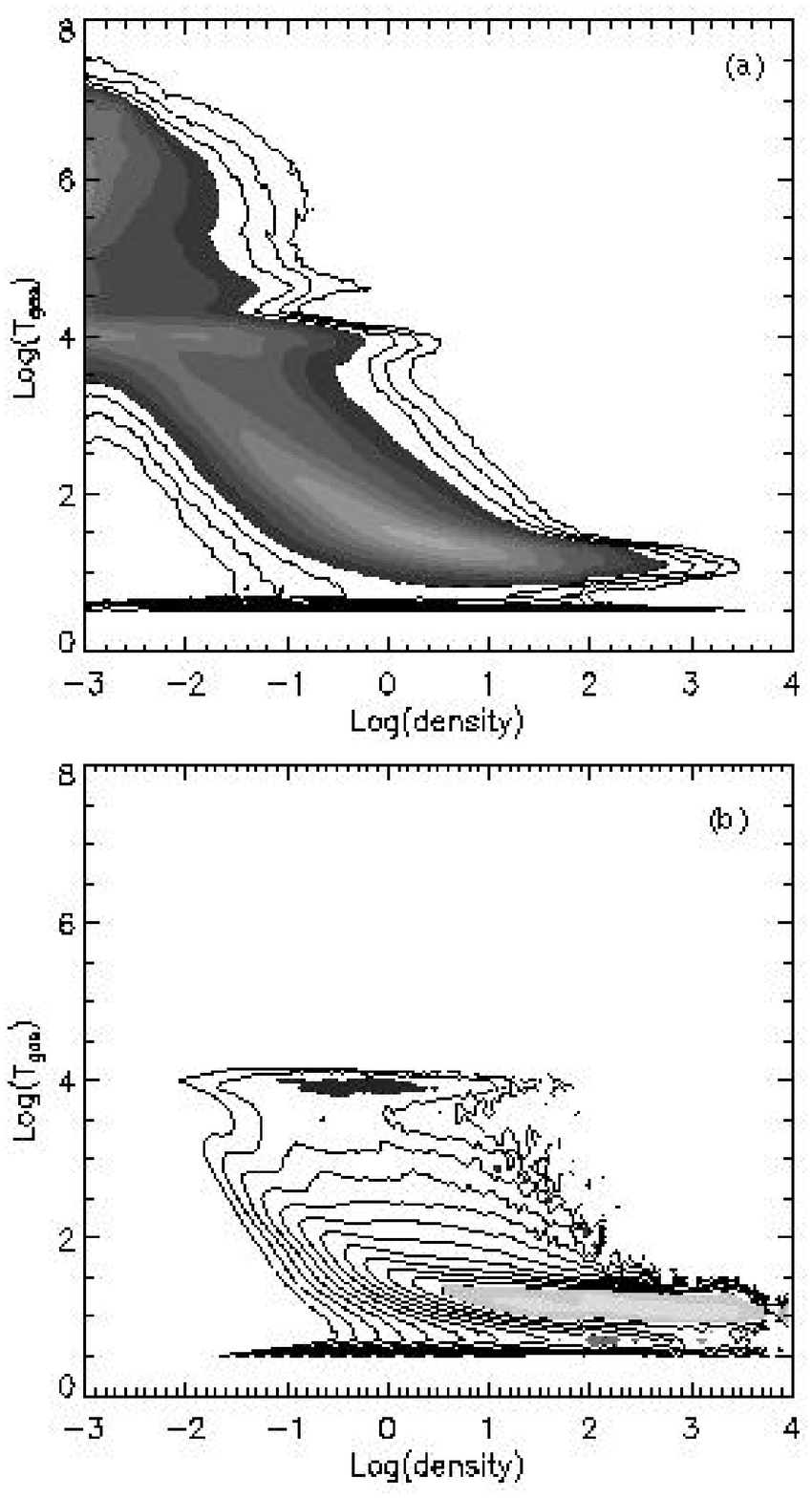}
\caption{(a) Volume-weighted phase diagram. 
The units of temperature and density are K and $M_\odot$ pc$^{-3} \sim 40$ H cm$^{-3}$. 
The 15 contour levels represent 
the number of grid-cells
in each phase (density and temperature), from $10^2$ to $10^6$. 
 (b) Same as (a), but for mass-weighted phase diagram.
The 15 contour levels represent mass in each phase (density and temperature), from $10^2 M_\odot$ to $10^6 M_\odot$. 
}
\label{wada_fig: f5}
\end{figure}

\begin{figure}[t]
\centering
\plotone{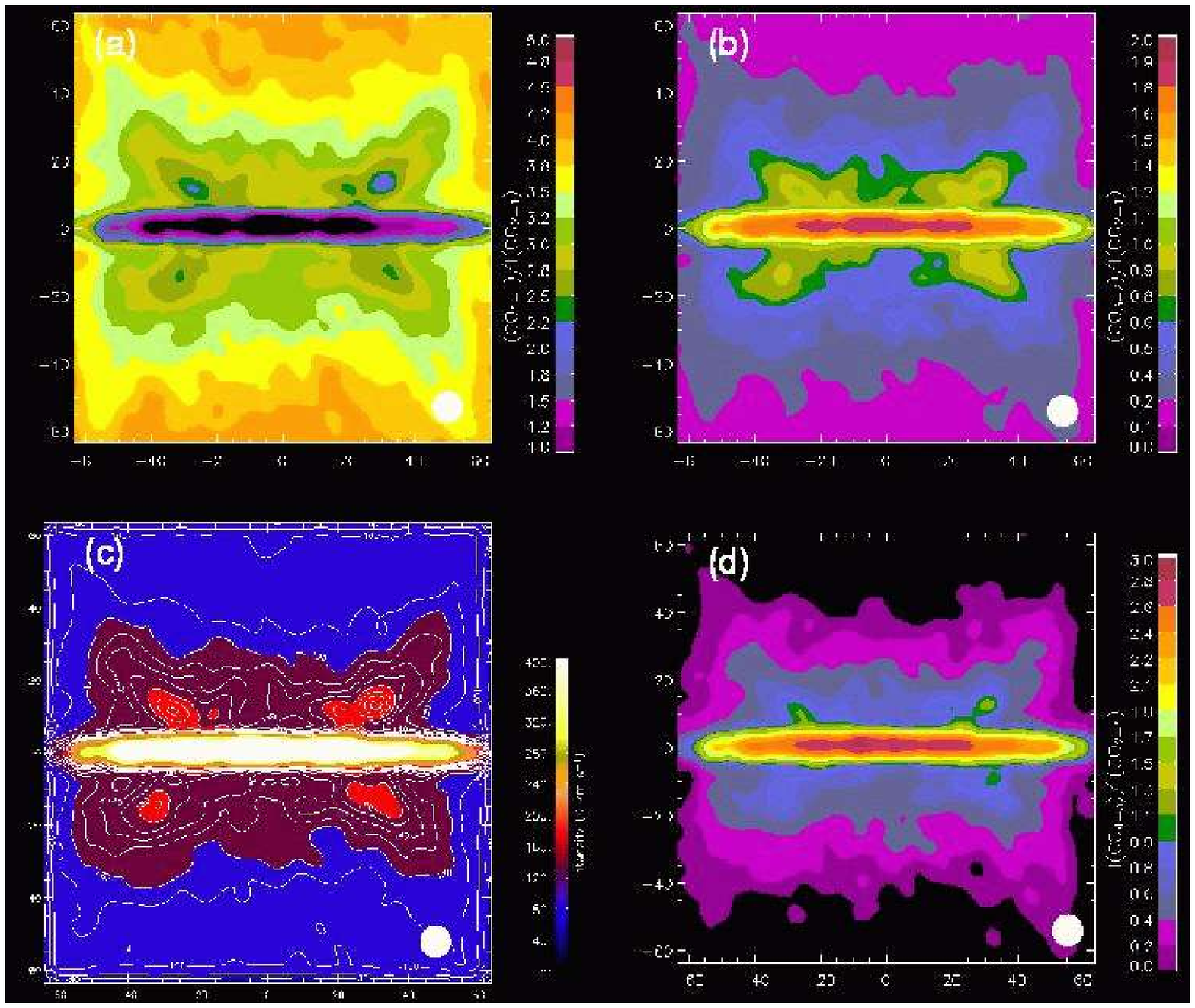}
\caption{$^{13}$CO integrated intensities of (a) $J=$1-0, (b) $J=$3-2, 
and (d) $J=$4-3 normalized for the intensity of $J=$ 2-1. The input model is model B and an edge-on view is assumed. Panel (c) is integrated intensity 
of $^{13}$CO (2-1) (unit is K km s$^{-1}$). 
Assumed beam-size is shown by a filled circle in each panel. Each box size is
64 pc $\times$ 64 pc. 
}
\label{wada_fig: f6}
\end{figure}

The two panels of Figure \ref{wada_fig: f6} are volume-weighted and mass-weighted 
histograms for a density and temperature in model B. As in the usual three-phases 
model of the ISM \citep{MO}, 
three stable branches can be seen in Figure 5a,
 i.e. cold, dense gases ($ 10 <\rho < 10^5 M_\odot$ pc$^{-3}$ and
$T_g < 10^2$ K),
warm gases at $T_g \sim 10^4$ K, and hot, diffuse gases at $T_g \sim 10^6$ K and
$\rho < 0.1 M_\odot$ pc$^{-3}$. 
The temperatures of warm and cold gases
are maintained by the photoelectric heating of the dust by the UV radiation, 
and shock heating in the compressible turbulence due to supernova explosions.
They are roughly in pressure equilibrium, but one should note
that the dispersion of the pressure is quite large for a given density or
temperature, which suggests that structure of 
the multi-phase ISM may not be determined under a constant pressure.
Figure \ref{wada_fig: f5}b, the mass-weighted phase diagram shows that 
the gas mass is dominated by the cold, dense phases. 
Mass fraction of the gas with $T_g < 50 $ K is about 20\% of the total mass.

Figure \ref{wada_fig: f5} shows line-ratios of 
$^{13}$CO for the edge-on view of model B.
The flared disk shown in 
Figure \ref{wada_fig: f2} is clearly
seen with the ratio of \thco(J=1-0)/\thco(J=2-1) $\sim 2$,
and \thco(J=3-2)/\thco(J=2-1) $\sim 1$,
 whereas in $^{12}$CO line-ratios (Figure \ref{wada_fig: f6}),
the flared disk is not clear.
These results suggest that the $^{12}$CO/$^{13}$CO line-ratio is not
uniform in the nuclear starbursts even for 
the constant abundance ratio that we assumed.
This kind of fine structure of the molecular gas should be detected
in nearby Seyfert 2 galaxies with starbursts \citep{stor01,cid03}
by future observations with the ALMA.

\begin{figure}[t]
\centering
\plotone{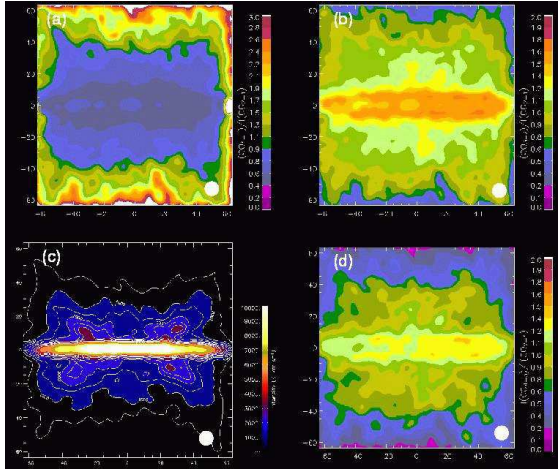}
\caption{Same as Figure \ref{wada_fig: f5}, but for \twco lines.
Panel (c) is \twco (2-1) integrated intensity distribution.
}
\label{wada_fig: f7}
\end{figure}

\begin{figure}[t]
\centering
\plotone{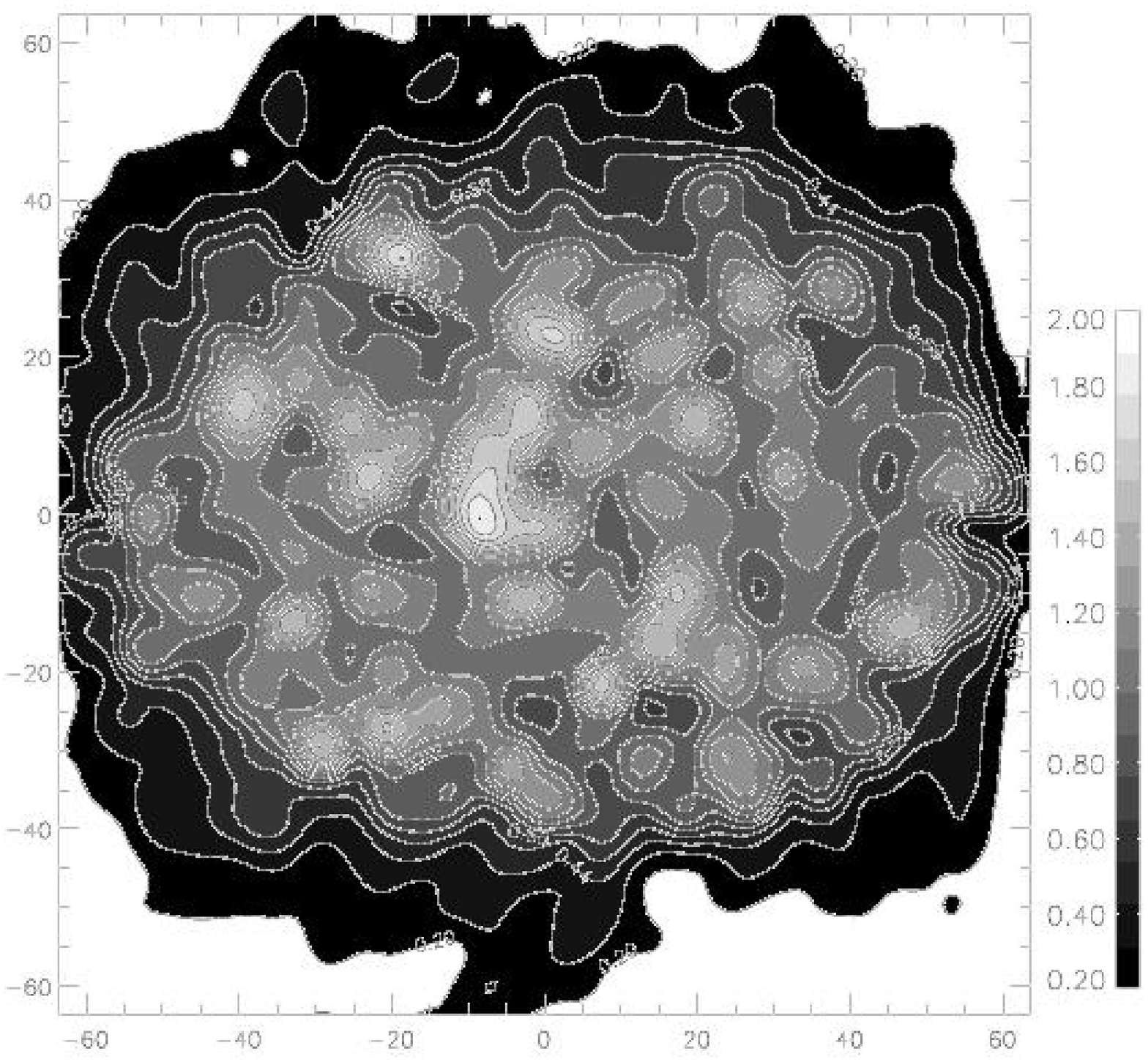}
\caption{Two-dimensional distribution of the CO-to-H$_2$ conversion factor, $X$ ($10^{20}$ cm$^{-2}$ [K km s$^{-1}$]$^{-1}$ , for \twco ($J=$1-0). Model B with a viewing angle of 45$^\circ$. Box size is 64 pc$\times$ 64 pc.}
\label{wada_fig: f8}
\end{figure}

Figure \ref{wada_fig: f8} shows distribution of the CO-to-H$_2$ conversion 
factor (X-factor) calculated from column density of the gas and
integrated intensity of \twco ($J=$1-0).
 The starburst model (Fig. \ref{wada_fig: f2}) is
observed from a viewing angle of 45$^\circ$, convolved with a Gaussian kernel
of FWHM 2.5 pc (five grid points). 
It is clear that the X-factor is not uniformly distributed.
In most regions, the X-factors
 are smaller than the value estimated for
the local clouds in the Galaxy [2-3 $\times 10^{20}$ cm$^{-2}$ (K km s$^{-1}$)$^{-1}$] \citep{solo87}.
Its distribution roughly follows the column density distribution.
This is also clear in Figure \ref{wada_fig: f7}, in which the X-factors
for four lines of $^{12}$CO are plotted against the integrated intensity. 
The X-factor of $^{12}$CO ($J=$1-0) depends on the intensity, and
the larger X-factor (i.e. higher density) is obtained for stronger 
intensities. In other words, the column density of the molecular gas in
dense regions may be under-estimated if a constant conversion factor is assumed.
On the other hand, the X-factor of $^{12}$CO ($J=$3-2) is nearly constant for an integrated intensity of less than 2000 K km s$^{-1}$.
These results suggest 
that $^{12}$CO ($J=$1-0) is not a good tracer for gas column density and 
total gas mass for the dense nuclear gas disk with the starburst, but 
$^{12}$CO ($J=$3-2) can be
used to estimate the column density of the molecular gas 
in a wide range of intensity. 
 The average X-factor for $^{12}$CO ($J=$3-2) is 
$(0.27\pm 0.04) \times 10^{20} {\rm cm}^{-2} ({\rm K} \;  {\rm km \;\; s}^{-1})^{-1}$
 in the range of $10^2 <I_{\rm CO} < 2\times 10^3$ K km s$^{-1}$.
These results are consistent with the well known fact that the mid-J CO
transitions are suitable for derivation of the column density of warm (e.g. 30-50 K)
 molecular clouds in our Galaxy\citep[e.g.][]{stu90}.
Using `synthesized observations' of the numerically obtained 
magnetized turbulent clouds, 
\citet{oss02} also showed that the X-factor for \twco (1-0) linearly
increases for the column density. 
They claimed that the X-factor is a constant
only in a small column density range, and this is consistent with
our results, although the objects (molecular clouds in a galactic disk
and a molecular torus around the AGN) are very different.

\begin{figure}[t]
\centering
\plotone{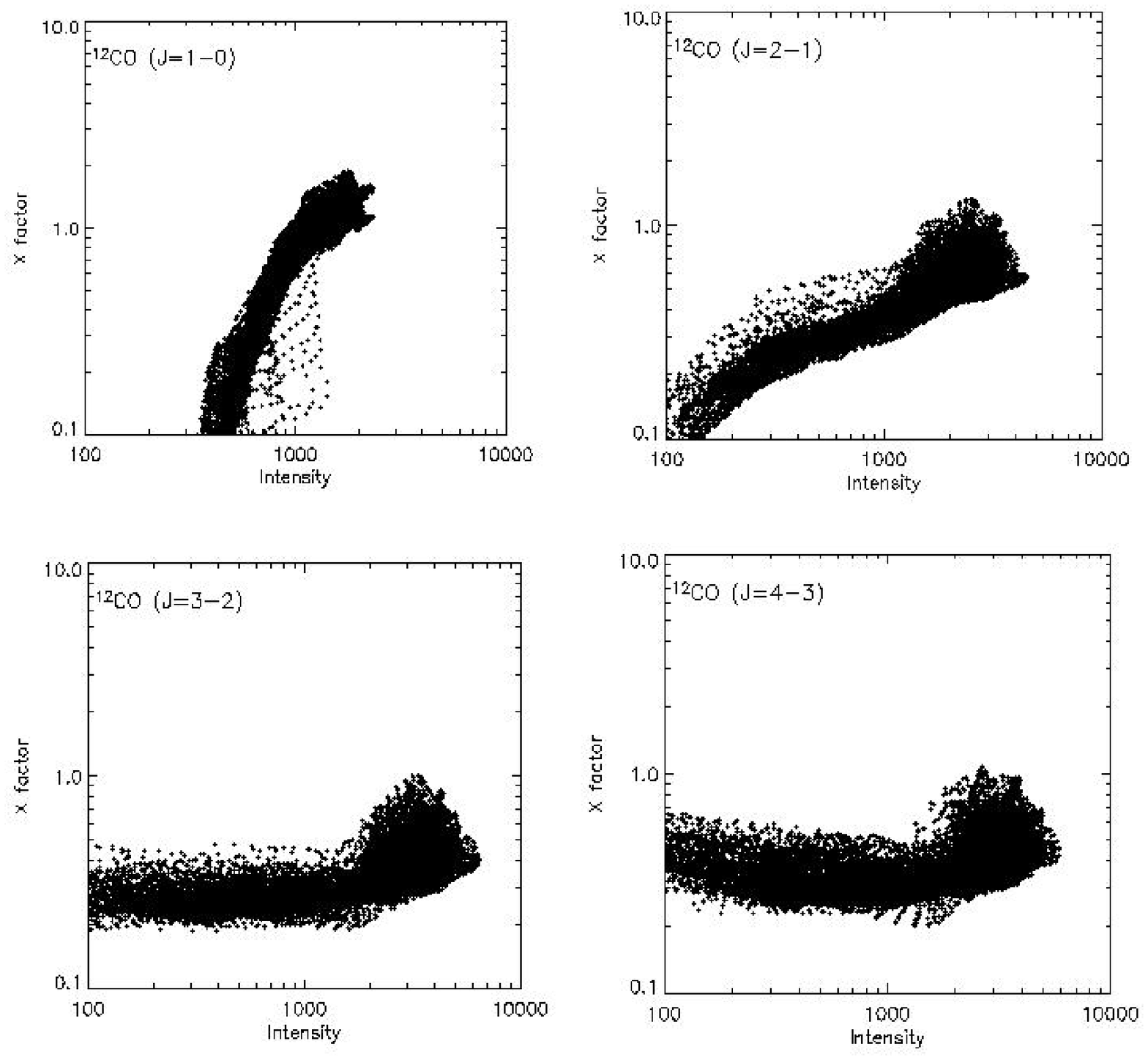}
\caption{CO-to-H$_2$ conversion factor, 
X [$10^{20}$ cm$^{-2}$ (K km s$^{-1}$)$^{-1}$] 
as a function of $^{12}$CO integrated intensity (K km s$^{-1}$) for
four transitions. Each dot represent a
grid point of the 2-D distribution of the X-factor like Fig \ref{wada_fig: f8}.
A viewing angle of 45$^\circ$ is assumed. }
\label{wada_fig: f9}
\end{figure}

\begin{figure}[t]
\centering
\plotone{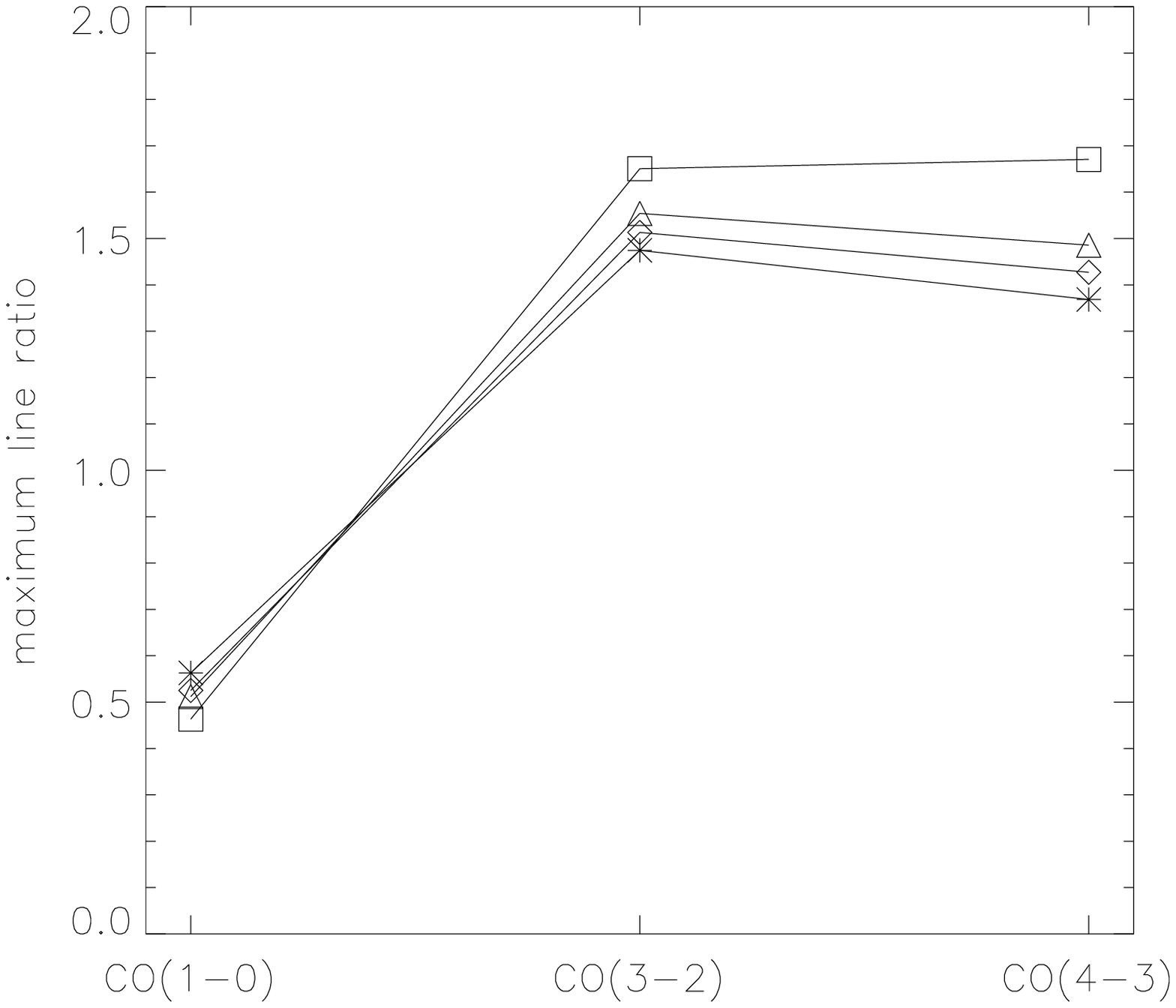}
\caption{Dependence of the maximum line-ratios to $^{12}$CO ($J=2-1$) on
the four ``beam sizes'', which are 1/8 (open boxes), 1/4 (open triangles), 1/2 (open
diamonds), and 1 (stars) radii of the disk. }
\label{wada_fig: f10}
\end{figure}

The line-ratios, such as $^{12}$CO (2-1)/$^{12}$CO (1-0), are
often used to infer the physical conditions of molecular clouds, mostly
in the Milky way (e.g. Oka et al. 1998), 
with the Large-Velocity-Gradient (LVG) technique \citep[e.g.][]{sco74}.
We find that the line-ratio is not also uniformly observed  for the nuclear disk. 
This suggests that the line-ratios observed with large beams do not necessarily show
the true ratios. In Figure \ref{wada_fig: f10},
the maximum line-ratios of $^{12}$CO (1-0)/(2-1), (3-2)/(2-1), (4-3)/(2-1) are plotted for four different beam sizes. One finds that the ratios depend on beam sizes, 
especially for higher-$J$ transitions. CO(4-3)/CO(2-1) is different by about 20\% between results with 
the finest and lowest resolutions. This implies that for a line-ratio obtained with
a beam size that is comparable to or larger than the object (i.e. nuclear disk),
the `true' line-ratio on a local scale could be larger than the average by
20\%. This difference might not be negligible in
inferring the physical conditions 
of the ISM using the usual LVG analyses. 
In this sense, we should understand that the currently observed molecular 
line-ratios in external galaxies (e.g. Wall et al. 1993; Zhu et al. 2003)
posses some intrinsic uncertainty.

The beam sizes of current radio interferometer observations
are typically a few arc-seconds, which is
comparable to or larger than the size of the molecular disk/torus 
in nearby galaxies. Therefore it is impossible to make a direct comparison
between our models and observations on the fine structure of the disk/torus. 
Here we compare our results 
with radio observations of central regions
of nearby galaxies on the peak intensity and the line ratio.
The Virgo high-resolution CO survey 
with the Nobeyama Millimeter-wave Array (NMA) by \citet{sof01} 
revealed that many normal spiral galaxies have molecular gas cores at 
their central several 100 pc regions,
and their \twco (1-0) intensity is typically 200-400 K km s$^{-1}$.
In Fig. \ref{wada_fig: f4}b, we showed that \twco (2-1) intensity is $\sim$ 3000 K km s$^{-1}$ 
on average,  and this is larger by one order of magnitude 
than the observed peak intensity of normal spiral galaxies 
Note that, as shown in Fig. \ref{wada_fig: f4}a, 
if the nuclear disks/tori have a pc-scale sub-structure,  
the maximum intensity could be much higher than 1000 K km s$^{-1}$.
By combining NMA and the 45m telescope at Nobeyama, 
\citet{sof01} and \citet{kod02} found a very dense molecular core in the central 
region of NGC 3079, which is Seyfert 2 with a nuclear starburst.
The \twco (1-0) peak intensity of this galaxy is extremely high,
$3.8\times10^3$ K km s$^{-1}$, and our torus model associated with the starburst
is consistent with this high peak intensity.

For higher-J transitions, such as CO (3-2) and (4-3), the beam sizes of 
current observations are typically $\sim$ 20 arc-seconds.
Therefore it is hard to make direct comparison between the observations and 
models.
Using the 15 m James Clerk Maxwell Telescope (JCMT), \citet{isr01, isr02, isr03}
presented $J=2-1, 3-2$ and 4-3 maps of \twco, and $J=2-1, 3-2$ maps of \thco
of 15 nearby spiral galaxies. The peak intensity of \twco (4-3) distributed
from 9 K km s$^{-1}$ (NGC 278) to 1019 K km s$^{-1}$ (NGC 253).
\citet{har99} showed that $I_{{\rm CO}(3-2)} = $998 K km s$^{-1}$ 
in the central 300 pc of NGC 253 using JCMT. They also showed 
that the line ratio, \twco (3-2)/\twco (2-1)$ = 1.1$ at the center of NGC 253.
Note that using the Heinrich-Hertz-Telescope, \citet{dum01} showed
$I_{{\rm CO}(3-2)} = $680 K km s$^{-1}$ and \twco (3-2)/\twco (2-1)$ = 0.7$.
The line ratio, \twco (3-2)/\twco (2-1), shown in Fig. \ref{wada_fig: f10} is $\sim 1.4$,
if the beam size is comparable to the disk radius.
Our results suggest that the line ratio in the central 100 pc
region is inhomogeneous, and the apparent ratio depends on the beam size.
To confirm this, we have to await future 
ultra-high resolution observations with the ALMA. 

%
\section{SUMMARY AND DISCUSSION}
%

In this paper, we made the first `observational visualization' of a 
molecular torus (or disk) around a supermassive black hole in an active 
galactic nucleus under the dynamical influence of starbursts through 
supernova explosions.  Three-dimensional hydrodynamic 
simulations of the multi-phase gas disk/torus are used as an input 
for the three-dimensional non-LTE radiative transfer calculations
for $^{12}$CO and $^{13}$CO lines. We found that clumpy and turbulent 
structure of the nuclear disk whose radius is about 30 pc can be resolved
with a few pc spatial resolution, which will be achieved with the ALMA for
nearby galaxies.  Our results suggest that the CO-to-H$_2$ conversion factor (X-factor) is not uniformly distributed in the cental 100 pc, and the X-factor for $^{12}$CO ($J=$1-0) is not constant against the density.
On the other hand, $^{12}$CO ($J=$3-2) is
nearly constant over the wide range, which is $\sim 0.3 \times 10^{20} {\rm cm}^{-2} ({\rm K} \;  {\rm km \; s}^{-1})^{-1}$, and it can be used to estimate
molecular gas mass in gas-rich galactic central regions 
associated with active star formation.
 We also found that the observed line-ratios depend
on the beam sizes, and the average line-ratios (e.g. \twco(4-3)/\twco(2-1))
do not necessarily represent the `real' ratios of the ISM on a local (parsec) scale.
We should be careful to use the ratios
to infer the physical conditions of the ISM if the spatial resolution is not 
fine enough. 

Although this is the first attempt to investigate 
the three-dimensional structure and 
dynamics of the ISM in the galactic central region, which could be compared directly
with observations by next-generation radio interferometers, 
our methodology still needs further improvement in various respects.
Firstly, we have not taken into account the radiation from the AGN itself.
The strong UV radiation from the accretion disk could affect both the dynamics and
excitation of the molecular gases (see also Ohsuga \& Umemura 2001). We also assumed a uniform UV radiation field. However this uniformity would not be the case in reality, because
the radiation from the massive stars distributed 
in the clumpy nuclear disk should affect the dissociation and 
excitation of molecules as well as the UV heating of the ISM non-uniformly. 
We cannot answer how serious these effects are on the `synthesized observations'
presented here, before we more consistently solve the radiation field. 

The CO abundance, which is assumed constant in our calculations, could
affect the intensity and the CO-to-H$_2$ conversion factor \citep{saka96}.
\citet{arim96} suggested that the observed CO intensities
depend on the metallicity in galaxies, and they found smaller X-factors in 
the central regions than in outer regions of galaxies. In this paper,
we assumed the solar metallicity, and a smaller X than the local value is 
suggested. Thus the X-factor could be even smaller in the nuclear starburst region
 with higher metallicity than the Galactic local value by a factor of two 
or maybe as much as ten.
In fact, low values of X have been suggested for the central regions of 
galaxies \citep[e.g.][]{wall93, reg00}.
It is certain at least that $^{12}$CO ($J=$1-0) with the standard X-factor inferred 
from observations for our Galaxy should not be used without caution, especially for
the dense ISM with strong star forming activities, such as starburst galaxies or
ultra-luminous infrared galaxies.

The dense molecular gas can be traced not only by CO lines, but also by ionized 
and atomic carbon, $[{\rm C_{II}}] 158 \mu m$ and by $[{\rm C_I}]$ lines. Under some conditions,
CO molecules are photo-dissociated by UV radiation, but H$_2$ is still present.
The neutral carbon may be ionized in the photon-dominated region (PDR).
Observations with the Kuiper Airborne Observatory revealed that 
$[{\rm C_{II}}]$ emission is correlated with the $^{12}$CO(1-0) emission
in gas-rich star forming galaxies \citep{sta91}.
The Balloon-borne Infrared Carbon Explorer (BICE) also found 
$[{\rm C_{II}}]$ emission in the galactic plane, which shows similar distribution to
CO \citep{nak98}. The Antarctic Submillimeter Telescope and Remote Observatory (AST/RO)
has recently revealed that $[{\rm C_{I}}]$ emission has a spatial extent 
similar to low-J CO lines in the Galactic central regions \citep{mar04}.
This is also the case in external galaxies, based on 
observations with the James Clerk Maxwell Telescope (JCMT)\citep{isr02}.
These observations support the PDR picture \citep[e.g.][]{tie85} in 
terms of the molecular clouds illuminated by the nearby OB stars.
In the present paper, we do not solve the UV radiation field 
in the torus, and photo-chemistry in terms of CO, $[{\rm C_{I}}]$, and $[{\rm C_{II}}]$ 
is not taken into account. One should note, therefore, that 
the CO intensities and the conversion factors shown in \S 4 
might be modified if the photo-chemistry and
the UV radiative transfer are fully taken into account.
Although such calculations are beyond the scope of the present paper, 
we can estimate how important the photo-chemical effects are using
the published PDR models.  \citet{moc00} investigated
PDR models of molecular clouds, solving chemical and thermal equilibrium 
with radiative transfer calculations for simple, static molecular clouds.
Their method is basically based on the PDR model by \citet{tie85}, but
they assume a spherical cloud instead of the plane-parallel geometry.
The molecular clouds considered in \citet{moc00} have typically a 
mass of $10-10^5 M_\odot$, a mean hydrogen density $10^2 < n < 10^4$, and
size 2-20 pc, and these are similar to the clumps in our torus models \citep{wad02}. Mochizuki \& Nakagawa (2000) found that 
$[{\rm C_{II}}]$ emission originates in the outer envelope of the
molecular cloud. On the other hand, CO emission is concentrated in
the inner region of the cloud. They also found that luminosity of CO
and the CO-to-H$_2$ conversion factor are insensitive for the far-UV flux
over about two orders of magnitude.
The conversion factor is close to the standard X-factor for a 
relatively massive cloud ($ > 10 M_\odot$). These results
imply that the UV radiation affects mainly the outer envelopes of
typical molecular clouds/cores; therefore, the CO intensity distribution
and conversion factor presented in the previous section would not
be significantly changed even if we took into account the photo-chemistry
in our model. Of course, this must ultimately be verified by 
fully self-consistent radiative transfer calculations with chemistry
for the clumpy, turbulent medium.


We showed in this paper 
that spatial resolution ten-times finer than present-day observations
in the radio frequency could drastically change our pictures of the ISM
 in the central regions of the external galaxies.  
Combining numerical techniques like the one presented here 
with the near-future observations by ALMA will be a very powerful tool for
investigating the unsolved problems associated with galactic nuclei, 
such as the starburst-AGN connection and fueling mechanism for the AGNs.

%
\acknowledgments 
%
We would like to thank the ALMA-Japan team, especially 
Tetsuo Hasegawa, Kotaro Kohno, Jin Koda, and Seiichi Sakamoto 
for helpful and stimulating discussions.
We are grateful to Nick Scoville for fruitful suggestions.
The anonymous referee gave us many important suggestions to improve the manuscript.
The authors also thank Ryoji Matsumoto, 
Yusuke Imaeda and Kazuya Saigo for their support and contribution through 
ACT-JST of the Japan Science and Technology Corporation.
Numerical computations were carried out on Fujitsu VPP5000 at NAOJ.  
KW is supported by Grant-in-Aids for Scientific Research 
[no. 15684003 (KW) and 14540233(KT)] of JSPS.

\end{document}